\documentclass[10pt,letterpaper]{article}

\usepackage{cogsci}
\usepackage[textsize = tiny,backgroundcolor=blue!10!white]{todonotes}

\cogscifinalcopy % Uncomment this line for the final submission 
\usepackage{graphicx}
\usepackage{caption}
\usepackage{subcaption}
\usepackage{xcolor}
\usepackage{placeins}
\usepackage{amsmath}
\usepackage{dirtytalk}
\usepackage{pslatex}
\usepackage{apacite}
\usepackage[symbol]{footmisc}
\usepackage{float}
\usepackage{multirow}
\usepackage{float}
\usepackage{tcolorbox}
\usepackage{placeins}
\definecolor{pink}{rgb}{0.92,0.87,0.89}
\title{\textit{Pencils to Pixels}:\\
A Systematic Study of Creative Drawings across Children, Adults and AI}

\author{{\large \bf Surabhi S Nath$^{1,2,3}$,  \large \bf  Guiomar del Cuvillo y Schröder$^{4}$, Claire E. Stevenson$^{4}$} \\
  $^1$Max Planck Institute for Biological Cybernetics, Tübingen, Germany\\
  $^2$Max Planck School of Cognition, Leipzig, Germany\\
  $^3$University of Tübingen, Tübingen, Germany\\
  $^4$University of Amsterdam, Amsterdam, Netherlands
  }

\begin{document}

\maketitle

\begin{abstract}
Can we derive computational metrics to quantify visual creativity in drawings across intelligent agents, while accounting for inherent differences in technical skill and style? To answer this, we curate a novel dataset consisting of 1338 drawings by children, adults and AI on a creative drawing task. We characterize two aspects of the drawings—(1) style and (2) content. For style, we define measures of ink density, ink distribution and number of elements. For content, we use expert-annotated categories to study conceptual diversity, and image and text embeddings to compute distance measures. We compare the style, content and creativity of children, adults and AI drawings and build simple models to predict expert and automated creativity scores. We find significant differences in style and content in the groups—children’s drawings had more components, AI drawings had greater ink density, and adult drawings revealed maximum conceptual diversity. Notably, we highlight a misalignment between creativity judgments obtained through expert and automated ratings and discuss its implications. Through these efforts, our work provides, to the best of our knowledge, the first framework for studying human and artificial creativity beyond the textual modality, and attempts to arrive at the domain-agnostic principles underlying creativity. Our data and scripts are available on GitHub\footnote{\textit{https://github.com/surabhisnath/pencils\_to\_pixels}}.

\end{abstract}

\section*{Introduction}

\begin{figure*}[ht!]
    \centering
    \includegraphics[width=\linewidth]{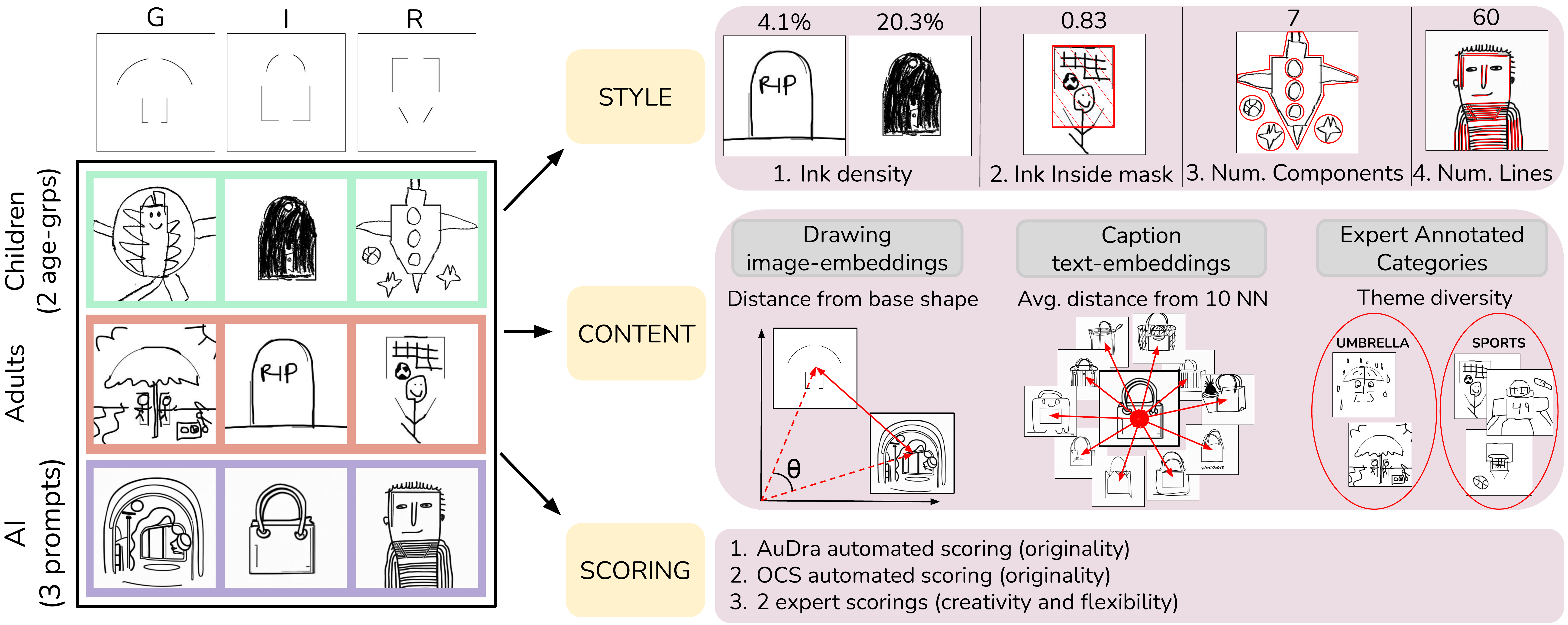}
    \caption{Overview of task and framework for studying creative drawing. The three MTCI stimuli G, I, R are shown on the top left below which example drawings by children, adults and AI are shown. Our style measures are shown on the top right panel. The middle right panel shows our content measures. The bottom right panel lists the different creativity scoring methods.}
    \label{figure1}
\end{figure*}

Human visual creative expression emerges early on---children start drawing before they can write \cite{levin2003drawingprecedes}, and cave paintings predate the written word \cite{ardila2004cavedrawings}. However, empirical research in creativity, 
especially in the context of AI, 
has focused more on verbal than visual creativity. This discrepancy arises in part due to the complexities of producing and evaluating visual outputs.
% Unlike the structured nature of verbal data, and the relative ease of evaluating textual creativity using predefined metrics,
Evaluating visual creativity involves subjective perceptual judgments, for example aesthetic considerations and challenges of separating creativity from technical skill \cite{chan2010drawingskill}. For this reason, visual creativity is commonly assessed using 
simple shape completion creative drawing tasks (e.g., TTCT Picture Completion, \cite{torrance1966TTCT}; TCT-DP, \cite{urban2005TCT-DP}; MTCI, \cite{barbot2018dynamics}). Unlike other forms of visual expression (e.g., paintings, digital art), creating such drawings requires limited technical or artistic expertise, without compromising on creative potential \cite{barbot2015domaingen} and serves as a useful cognitive tool \cite{fan2023drawing}. 

However, evaluating such drawings are still challenging due to the lack of a well-formalized framework for assessment, especially across intelligent agents, a crucial comparison in the present era where human and machine creativity increasingly intersect \cite{acar2023creativity, o2024extending, marr2023intersection}. For example, how can we meaningfully compare the creativity of a child's pencil drawing to that of a $1024 \times 1024$ pixel image generated by an AI model? To bridge this gap, we make two key contributions: (1) we curate a novel dataset of creative drawings and evaluations spanning different intelligent agents, (2) we develop a computational framework that quantifies two core aspects of the drawings---content, and style, and use them to study drawing creativity.
% and arrive at ``universal" computational metrics of visual creativity that characterize both style and content, so that we can compare visual creativity
% across intelligent systems.

\noindent\textbf{Dataset.} In order to create a diverse dataset, we take note of the vast literature studying individual differences in drawing abilities \cite{chan2010drawingskill}, particularly across development \cite{lowenfeld1957drawingstages, philippsen2022quantifying, heard1988children, hart2022development, narvaez2024children}. Further, we note that in AI text-to-image models, prompting can lead to significant diversity in outputs \cite{oppenlaender2024prompting}. Therefore, we curate drawings from children in two age groups, adults, and AI models prompted using a collection of prompts. 

For each drawing, we obtain human ratings from two experts raters and get automated scores from two recently released tools for automated assessment of drawing creativity, AuDrA \cite{patterson2024audra} and OSC-figural \cite{acar2023autofigural}. 
% For evaluating the drawings, it is important to consider potential biases in assessment methods. 
Research suggests that AI-based evaluation methods favor AI-generated responses, whereas human evaluators prefer human-created outputs \cite{laurito2024ai}. Therefore, by incorporating both expert \cite{kaufman2012beyond} and automated evaluation \cite{cropley2022automated}, we obtain a balanced perspective on creativity assessment.
% This multi-perspective approach enables a comprehensive analysis of how productions and evaluations by different intelligent agents compare.

% For each drawing, we obtain human ratings from two ex-
% perts raters and get automated scores from two recently re-
% leased tools for automated assessment of drawing creativity,
% AuDRA (Patterson, Barbot, Lloyd-Cox, & Beaty, 2024) and
% OSC-figural (Acar, Organisciak, & Dumas, 2023), both re-
% porting high correlations with expert scores. For evaluating
% the drawings, it is important to consider potential biases in
% assessment methods. Prior research suggests that AI-based
% evaluation methods favor AI-generated responses, whereas
% human evaluators prefer human-created outputs (Laurito et
% al., 2024). Therefore, we incorporate both human and AI-
% based evaluation frameworks to obtain a balanced perspective
% on creativity assessment.

\noindent\textbf{Framework.} To analyze this diverse dataset, we begin by distinguishing two key aspects of creative drawings: content (\textit{what} is depicted) and style (\textit{how} it is rendered). This distinction between content and style is studied in vision research, for example in the context of art 
% research to identify the time-course of
perception \cite{augustin2008style, augustin2011neural} and in generative artificial intelligence
% to facilitate image-style transfer 
\cite{kotovenko2019content, zhang2018separating}. 

We explore various computational metrics for characterizing content and style in creative drawings. For style, the number, length, distribution or smoothness of strokes are popular choices \cite{thomas2022identifying}. Other measures related to order or complexity from aesthetics research can be useful candidates \cite{nath2024relating, van2020order}.
%For automated style metrics, Nath et al. \citeyear{nath2024relating} found that the number of visual components and local spatial complexity could predict how aesthetically pleasing pixel-based images were. 
% Thus candidate computational metrics to characterise creative drawing style could be ink density, ink distribution in- and outside the stimulus lines and the number of visual elements (see top right of Figure \ref{figure1}). 

% It has also been applied in non-visual domains such as 
% % for disentangling writing styles vs topics in 
% text \cite{wegmann2022same, boghrati2023style}. Recent work in textual creativity studied story generation in humans and LLMs and concluded that creativity in storytelling involves the interplay of content novelty and linguistic style, measured using computational measures such as syntactic complexity and semantic coherence \cite{ismayilzada2024evaluating, marco2024small, ramesh2022automated}.

For content, we propose a multimodal approach that utilizes the drawing itself, captions generated for the drawing, or expert annotations. For this, one could begin by extending the measures developed for textual responses \cite{ismayilzada2024evaluating, ramesh2022automated}. For example, \textit{SemDis}, measuring the distance between a response and the task stimuli \cite{beaty2021semdis}, or inverse frequency, quantifying the uniqueness of a response \cite{weaver2019interpreting}, or process flexibility, tracking jumps in the responses sequence \cite{nath2024creaprocess} can be adapted for drawings.

Together, with this dataset and framework, we are the first to systematically characterize differences in creative expression across intelligent systems, and arrive at core computations underlying visual creativity, and creativity generally.
% , or creativity more generally.
% For this, we address the following research questions: \textit{How do drawings vary along style, content and overall creativity across ?},  

% Our main contributions are: curating a novel dataset of 1500 drawings stemming from children, adults and AI; devising computational measures of style and content in visual creativity; providing insights on differences in creative drawing between intelligent systems; and providing insights on which aspects of creative drawings are valued by expert versus automated raters. 

\section{Methods}
\vspace{-1mm}
\subsection{Data}
Our dataset consists of 1338 drawings by children, adults and AI on stimuli from the Multi-trial Creative Ideation (MTCI) Task \cite{barbot2018dynamics} (Fig~\ref{figure1} left panel).

\subsubsection{Children}
Data comprises of 444 drawings from 148 children, 84 from kindergarten (4-6 year olds) and 64 from lower elementary level (7-9 year olds) from a public Montessori school. These two groups align with two stages of drawing development, where ages 4-6 are considered \textit{pre-schematic} and ages 7-9 \textit{schematic} \cite{lowenfeld1957drawingstages}. 

Data collection took place in the classroom in small groups. A trained research assistant explained the task. Then each child was given a thick pencil and a piece of paper with the stimulus printed on it. They were given 5 minutes to complete their drawing, after which the next stimulus was given. Each child completed three drawings, one for each of the stimuli shapes G, I and R. 

Since the children drew freely on paper, they sometimes ignored instructions and flipped the paper by $180^{\circ}$, drawing on the inverted stimulus. Stimulus `R' (resembling an inverted house) was flipped most often, with 60\% drawings on `R' being inverted. Across the whole dataset, about 30\% drawings (nearly the same ratio in both pre-schematic and schematic) were drawn on inverted stimuli.

\subsubsection{Adults}
Data comprises of 444 drawings from 148 participants, who each completed drawings for stimuli shapes G, I, and R on the MTCI hosted by Barbot's Crealyx platform, an open-access, online testing platform dedicated to the assessment of creativity.

\subsubsection{AI}
We treated the MTCI task as an inpainting task. We prompted Open AI's Dall-e (v2), through their API in image editing mode using three prompts (see box below) for each of three stimuli shapes G, I, R. We collect a total of 50 images per prompt per stimuli, resulting in a final dataset of $(50 + 50 + 50) \times 3 = 450$ images.
%With the prompts, we wanted to test for (1) out of the box zero-shot creativity of Dall-e and (2) the potential for prompt-guided image content manipulation. 

The first prompt, based on Chen \citeyear{Stevenson_Chen_2023} contained clear stylistic instructions (colour, pen type, thickness, art style etc.) with no explicit instruction for content (to match the task instructions given to humans), and was set as the base prompt (prompt 1). 
Prompts 2 and 3 extended the base prompt with explicit content instructions for creating objects/scenes (prompt 2) and living figures (prompt 3). 

% The base prompt was borrowed from Stevenson and Chen \citeyear{Stevenson_Chen_2023}---they conducted a Monte Carlo experiment on Dall-e to identify prompts that generated valid outputs (\textit{i.e.} uncensored responses that could reasonably be created by humans within the scope of MTCI). The three prompts we used were as follows:
% Our resulting prompts were the following:

% Two-thirds of the generated drawings did not adhere to the instructions, (ignored the stimulus or added colours) so we generated 1350 drawings to yield a dataset of 450 drawings.

\begin{tcolorbox}[ colback=pink!10, colframe=pink!80!black, title={Prompts}]
\small
\textbf{Prompt 1}: \textit{creative minimalist black-on-white drawing, lineart-style on white background, drawn with digital fineliner, no color or shading}
\\
\\
\textbf{Prompt 2}: \textit{creative minimalist black-on-white drawing of day-to-day object or scene, lineart-style on white background, drawn with digital fineliner, no color or shading}
\\
\\
\textbf{Prompt 3}: \textit{creative minimalist black-on-white drawing of living figures (human, animal or creature), lineart-style on white background, drawn with digital fineliner, no color or shading}
\end{tcolorbox}

% Since two-thirds of the generated drawings did not follow the instructions (either ignored the stimulus or added colors), we produced a total of 1350 drawings to obtain a final dataset of 450 valid ones.

Two-thirds of the generated drawings did not follow the instructions (either ignored the stimulus or added colors), so we produced 1350 drawings to obtain 450 valid ones.

% \begin{enumerate}
%     \item Prompt 1: \textit{creative minimalist black-on-white drawing, lineart-style on white background, drawn with digital fineliner, no color or shading}
%     \item Prompt 2: \textit{creative minimalist black-on-white drawing of day-to-day object or scene, lineart-style on white background, drawn with digital fineliner, no color or shading}
%     \item Prompt 3: \textit{creative minimalist black-on-white drawing of living figures (human, animal or creature), lineart-style on white background, drawn with digital fineliner, no color or shading}
% \end{enumerate}

% The total amount across all experiments was $\sim$40\$. 
\subsection{Preprocessing}
Since the children, adults and AI data came from different sources, it was crucial to preprocess the images for valid comparisons. We controlled for size, colours and line thickness using computer vision techniques. The children and Dall-e drawings are cropped and resized to $400\times400$ size to match the adults. The cropping was specified in a way to ensure the stimulus size and position was aligned across all drawings. The children's drawings were made with pencil and are therefore shades of gray. All drawings were binarized and cast to black-and-white. Children drawings also had tiny scattered pencil spots which were removed using image erosion. 
%Average line thickness per drawing were estimated using image skeletonization to extract the structural outline followed by applying the distance transform. 
The lines for children and AI drawings were dilated to match the line thicknesses of adult drawings. Post processing, we confirmed a non-significant difference in line thickness across the three groups using a Kruskal-Wallis test ($p > 0.1$).

% example where same content but different creativity scores, same style, different content 
% domain general (content) vs domain specific (style)

\subsection{Measures}
We developed computational measures of style and content to characterise drawing creativity.

\subsubsection{Style}
We quantify (1) ink density, (2) fraction of ink inside the stimuli shape, (3) number of components, and (4) number of lines.

(1) \underline{Ink density}: The percentage of the drawing covered in ink, measured by dividing the number of black pixels by the total number of pixels times 100 (Figure \ref{figure1} Style panel 1.).

(2) \underline{Fraction\textcolor{white}{g}of Ink inside the Stimulus Shape}: Quantified by the amount of ink inside the stimulus's bounding box divided by the total amount of ink. Bounding boxes for the base stimuli were obtained by identifying the extreme ink points defining their boundaries. (Figure \ref{figure1} Style panel 2.).

(3) \underline{Number of Components}: Counts the number of visually distinct regions in the drawing, 
%Inspired by \cite{nath2024relating}, this measures counts the number of connected components in the drawing where a component is defined 
based on the graph theoretic property of reachability (based on \cite{nath2024relating}, Figure \ref{figure1} Style panel 3.). 
% \textit{I.e.}, a black pixel and all the black pixels you can reach from it (if you only move along black pixels without crossing any white pixels), are part of the same component.

(4) \underline{Number\textcolor{white}{g}of Lines}: Skeletonizes the drawing to extract its structural outline and then applies the Hough Transform to detect and count straight lines (Figure \ref{figure1} Style panel 4.).

\subsubsection{Content}
We use the \texttt{clip} image-embedding model (OpenAI, \texttt{clip-vit-large-patch14}) to obtain image embeddings per drawing. We compute the cosine distance between the embeddings of the drawing and the corresponding base stimulus shape (G, I or R).
% , serving as a measure of originality. 
Based on the nature of \texttt{clip}'s training data, we know the distance measures induced by these encoders are largely influenced by semantic rather than stylistic attributes \cite{udandarao2023visual, rashtchian2023substance}.

For captions, we use \texttt{GPT4o} to generate image descriptions for each drawing. We explicitly state in the prompt to give short captions ($<$15 words), describing the content (and not style) of the drawing, and to caption ``hard to interpret" drawings as such. We then used \texttt{gtelarge} text-embedding model to obtain caption-embeddings per caption per drawing. Using these, we (1) cluster the embeddings hierarchically to arrive at core semantic themes expressed in the drawings. (2) We define a measure of semantic uniqueness for each drawing by computing the mean cosine distance of the caption-embedding to its ten nearest neighbours. The nearer a caption-embedding is to others, the more popular and less unique the drawing concept is in the dataset of drawings. 
% This measure is analogous to inverse frequency and serves as another measure of originality.

\subsection{Annotation}
Drawing content was also annotated by an expert. Each drawing was classified into a minimum of one and a maximum of three (from most salient to least salient) concept categories. If the drawing was hard to interpret, it was assigned to a ``hard to interpret" category. Using these categories, we evaluate conceptual diversity for the different groups by diving the number of unique categories per group by the total number of unique categories. 

Drawing captions generated by \texttt{GPT4o} were also validated by an expert with a correct/incorrect label per caption per drawing. We found that \texttt{GPT4o} captions scored nearly 83\% correct across all drawings.

To measure process aspects, we obtained flexibility (conceptual diversity within the outputs of the same agent) scores by two expert raters. Flexibility was scored in a range of [0-2] for the three drawings by the same participant. Since there are no explicit participants for the Dall-e drawings, we randomly sample sets of three Dall-e drawings, one from each stimulus shape under the same prompt. We sample 50 random samples sets per prompt (without replacement so that each image is sampled once) and obtain flexibility scores for each set.

\subsection{Creativity Scoring}
Creativity was scored by four sources. Two experts rated each drawing on a scale of [0-4] following the MTCI scoring protocol.
%We also leveraged two existing automated scoring tools, specifically trained for MTCI, namely 
Two automated scoring tools AuDrA \cite{patterson2024audra} and Open Creativity Scoring - Figural \cite{acar2023autofigural} produced originality ratings in the range [0-1] per drawing. 

One expert rater also scored utility in the range [0-2] per drawing, depending on how well the drawing incorporated the base shape.
%Using the methods described in this section (also summarised in Figure \ref{figure1}), we study three main questions: (1) how do children, adult and AI drawings differ? (2) which group is more creative? and (3) what do human experts and automated tools value when rating visual creativity? We present our results in the following section.

\section{Results}

\subsection{1. How do Children, Adults and AI Drawings Differ?}
We compare the drawings of children, adults and AI based on style and content. We present the subgroups of children and AI separately to note differences across childhood development and prompt-guided content manipulation.

\subsubsection{Style} 
\begin{figure}
    \centering
    \includegraphics[width=\linewidth]{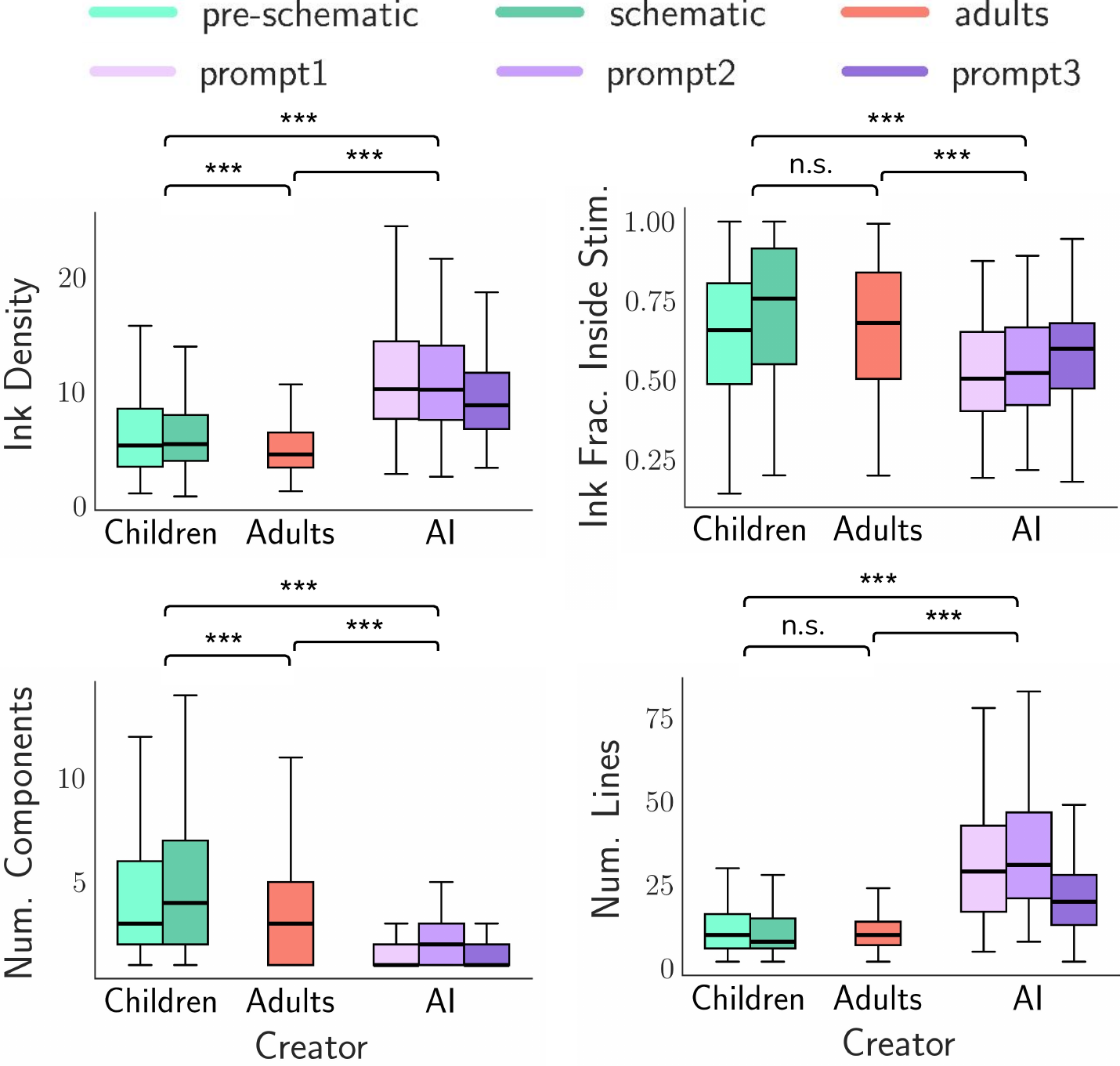}
    \caption{Comparing style measures of ink density, ink fraction inside stimulus, number of components and number of lines in drawings per subgroup. *** denotes $p < 0.01$.}
    \label{figure2}
\end{figure}

Figure \ref{figure2} presents the boxplots comparing the style measures. We see significant differences in ink densities across the three groups (the within group differences were not significant, Figure \ref{figure2}a). We find that AI drawings have the highest ink density followed by children and then adults with the least. From Figure \ref{figure2}b we see that most of this density resides inside the stimulus shape for children and adults with no significant difference between them
(there was a significant difference within the subgroups of children),
but resides outside for AI drawings. Further, the children's  drawings had a higher number of components compared to adults, and AI had the least (Figure \ref{figure2}c). This means the children draw more separable elements in their drawings whereas the AI produces a connected chunk of ink.
% (with no significant difference within the AI subgroups)
Finally, from Figure \ref{figure2}d we see that AI drawings contain a large number of straight lines, significantly higher than children or adult drawings (with no significant difference between children and adults groups or subgroups). Within AI, prompt 3 drawings had significantly lesser lines than prompt 1 or 2, suggesting that the instruction of producing living figures resulted in more curved strokes. Lastly, we found that while children and adults nearly always seamlessly incorporated the stimulus shape in their drawings, AI only did so about $50\%$ of the times.

\begin{figure*}
    \centering
    \includegraphics[width=\linewidth]{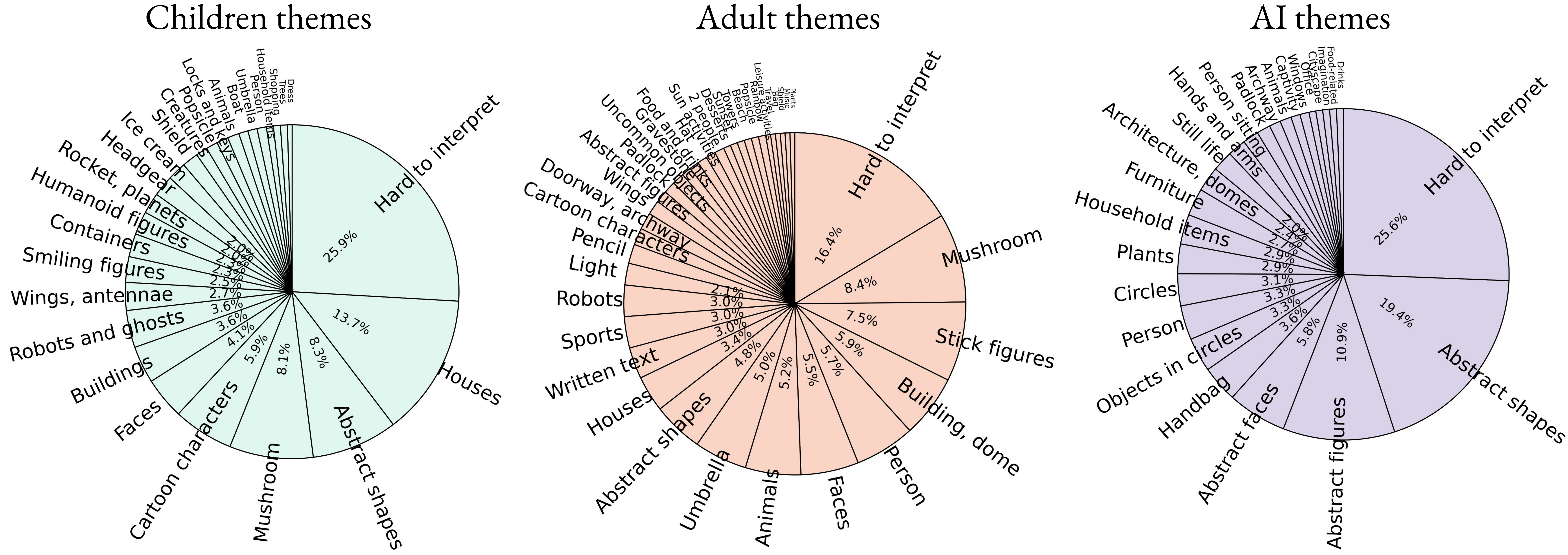}
    \caption{Conceptual themes present in children, adult and AI drawings.}
    \label{figure3}
\end{figure*}
    
\subsubsection{Content}
To study the content of the drawings, we identify the common themes across drawings using two methods. 

We use expert annotated categories to calculate concept diversity as the number of unique categories per subgroup divided by the total number of unique categories across subgroups (counted to be 253). Adult drawings displayed the maximum content diversity, with their drawings encompassing nearly 50\% of all identified categories, followed by children ($\sim30\%$) and then AI ($\sim18\%$) (wherein prompt 2 had a significantly higher diversity than other prompts). In the AI drawings, prompt 2 produced more object themes and prompt 3 produced more living themes, compared to prompt 1, suggesting that the prompting was effective. 

To understand the classes of themes more closely, we visualise them by hierarchical clustering of \texttt{GPT4o} caption-embeddings (Figure \ref{figure3}). We see that adults display the maximum number of themes, some of them also encompassing complex ideas e.g. gravestone, music or leisure activities. The children and AI drawings had a roughly similar number of categories but the themes differed significantly. Children themes included many imaginative concepts such as cartoon characters, wings, antennae, rockets, ice cream or ghosts. The large proportion of houses can be attributed to 60\% of `R' stimuli drawings being inverted. Many of the AI drawings contained abstract themes such as abstract shapes, figures or faces, while some others were complex themes such as furniture or captivity. Children and AI had a relatively similar fraction of drawings which were hard to interpret ($\sim25\%$), higher than those in adults ($16\%$). 

We plot the mean flexibility scores for each group to study process level differences (Figure 4). We note are adults are generally highly flexible with mean flexibility scores largely above 1. Within children, the pre-schematic group has relatively low flexibility with many children scoring a mean under 1. This means these children redrew the same themes across their drawings, disregarding the stimulus. However, in the schematic group, children are strikingly more flexible ($p < 0.01$), rarely repeating themes across their drawings. A similar shift towards higher flexibility is visible as a result of prompting. The base prompt yielded inflexible drawings, whereas prompt 2 and 3 sees a rightward shift in the values, yielding a significant difference in case of prompt 2 (also in line with the greater conceptual diversity in prompt 2).

\begin{figure}[h]
    \centering
    \includegraphics[width=0.9\linewidth]{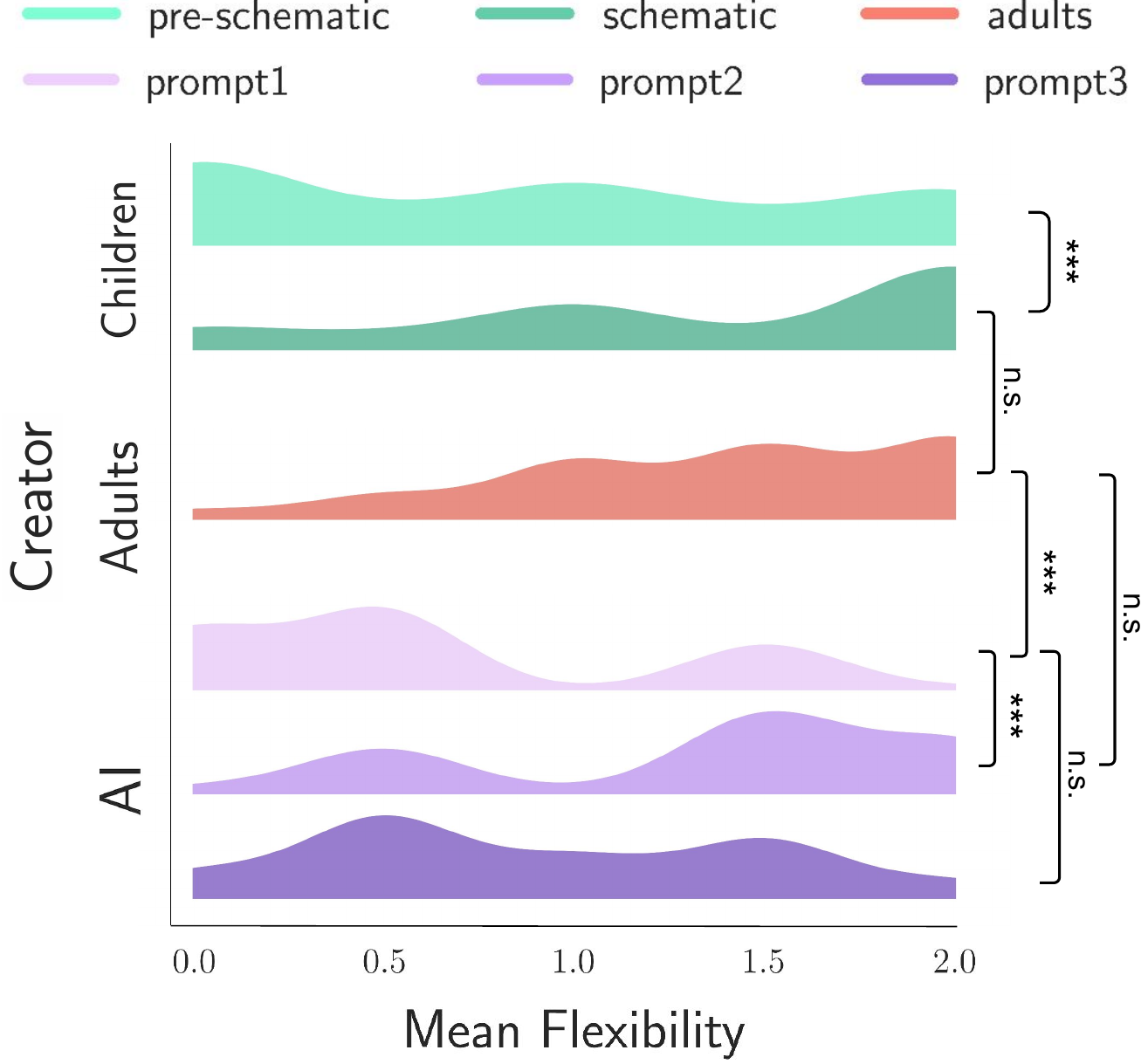}
    \vspace{0.3cm}
    \caption{Comparing drawing flexibility across subgroups. *** denotes $p < 0.01$.}
    \label{figure4}
\end{figure}

% \begin{figure}
%     \centering
%     \includegraphics[width=0.6\linewidth]{figures/Figure3_diversity.pdf}
%     \caption{Caption}
%     \label{fig:enter-label}
% \end{figure}

% \subsubsection{VisDiff}
% Done
% \subsubsection{Process Aspects}

% \subsubsection{Classifier}
% \textcolor{red}{TODO: We build a classifier to test how easily the groups and subgroups can be distinguished from each other based on the measures we develop.}

\subsection{2. Who is More Creative?}

To test which group received the highest creativity scores, we first check the agreement within and between the expert and automated scores. Between the two expert raters, and between the two automated methods, the average fixed raters ICC score was respectively 0.82 and 0.90 ($p < 0.01$), indicating high inter-rater reliability and consistency, and therefore we use mean expert score (normalised using min-max scaling to the range 0-1) and mean automated score for analyses. Interestingly, the ICC between mean expert score and mean automated score is only 0.48 and Spearman correlation coefficient is 0.36 ($p < 0.01$). This indicates that the experts and automated scores, though both reliable, do not agree with each other fully on how creative the drawings are. 

\begin{table}
    \centering
    \small
    \begin{tabular}{c|c|cc}
         & &  \textbf{Mean score} & \textbf{Mean score}\\
         & &  \textbf{expert} & \textbf{automated}\\
         \hline
         \hline
         \multirow{2}{*}{Children} & Pre-schematic & 0.38 ($\pm$ 0.26) & 0.50 ($\pm$ 0.12) \\  
          & Schematic & \textbf{0.50} ($\pm$ 0.24) & 0.53 ($\pm$ 0.11) \\  
         % & & & \\  
         Adults & & 0.43 ($\pm$ 0.25) & 0.49 ($\pm$ 0.12) \\  
         % & & & \\
          & Prompt1 & 0.21 ($\pm$ 0.16) & 0.57 ($\pm$ 0.08)\\  
         AI & Prompt2 & 0.36 ($\pm$ 0.17) & \textbf{0.62} ($\pm$ 0.08)\\  
          & Prompt3 & 0.32 ($\pm$ 0.18) & 0.61 ($\pm$ 0.07) \\  
    \end{tabular}
    \caption{Mean and standard deviation of drawing creativity scores by expert raters and automated tools per subgroup. Bold incates the subgroup with the highest score.}
    \label{table1}
\end{table}

\begin{figure}
    \centering
    \includegraphics[width=0.9685\linewidth]{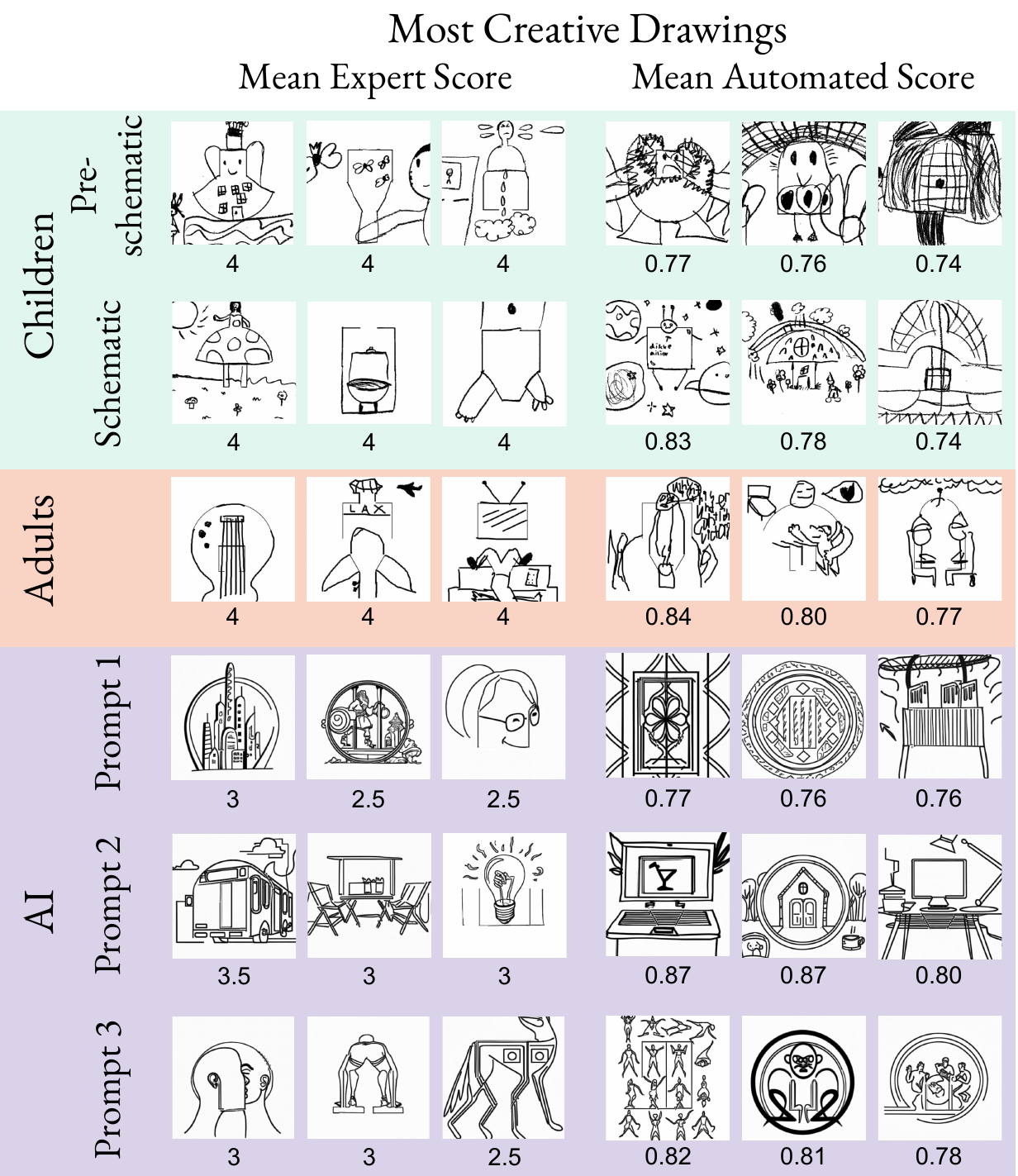}
    \caption{Top three drawings per subgroup receiving the highest mean creativity scores by expert raters, and automated tools. Value indicates mean creativity score.}
    \label{figure5}
\end{figure} 

\begin{table*}
    \centering
    \begin{subtable}{\textwidth}
        \centering
        \begin{tabular}{l|cc}
             Model & $R^2_{test}$ & $Cor_{test}$\\
             \hline
             \hline
             expert \textcolor{white}{llllll}$\sim$ (used\_stim + hard\_to\_interpret) * (10NN\_text + dist\_from\_stim) + (1 $\mid$ subgroup) & 0.60 & 0.79\\
             automated $\sim$ ink\_density + dist\_from\_stim + 10NN\_image + (1 $\mid$ subgroup) & 0.59 & 0.78\\
        \end{tabular}
        \caption{Best models predicting expert and automated scores.}
        \label{table2a}
    \end{subtable}
    \hfill
    \\
    \vspace{2mm}
    \begin{subtable}{\textwidth}
        \centering
        \begin{tabular}{l|cccccc}
              & used\_stim & hard\_to\_interpret & 10NN\_text & dist\_from\_stim & ink\_density & 10NN\_image\\
             \hline
             \hline
             expert\textcolor{white}{mmmllll} & \textbf{0.16}* & \textbf{-0.77}* & \textbf{0.14}* & \textbf{0.35}* & 0.15* & -0.04\textcolor{white}{*}\\
             automated & 0.00\textcolor{white}{*} & 0.03\textcolor{white}{*} & 0.07* & \textbf{0.34}* & \textbf{0.44}* & \textbf{0.10}*
        \end{tabular}
        \caption{Feature contributions for the two best models. * indicates $p < 0.01$. Bold marks the higher weight across the two models.}
        \label{table2b}
    \end{subtable}
    \caption{Comparison of model performance and feature contributions}
    \label{table2}
\end{table*}

Further, we test the agreement within the groups to find that the Spearman correlation between mean expert score and mean automated score is higher in adults (0.66, $p<0.01$), compared to children (0.59, $p<0.01$, with 0.56 in pre-schematic), and lowest in AI ($\sim0.30$, $p<0.01$ for each prompt subgroup). This can be explained by the fact that the automated models were trained on adult drawings and can therefore not generalize easily to other subdomains such as drawings by a different group of humans like pre-schematic children, or to non-human, artificial drawings by AI.

Table \ref{table1} reports the mean and standard deviation of mean creativity score by experts and automated tools for each subgroup. We see that the experts rated the schematic group as most creative, followed by adults and rated AI drawings the least creative. On the other hand, the automated tools rated adults the least creative, followed by children and scored the AI drawings as most creative. Across expert and automated scoring, within children's drawings, the schematic group scored higher than the pre-schematic group, and within AI drawings, prompt 2 scored the highest, followed by prompt 3 and the base prompt was scored the lowest. 
% TODO: Within human drawings, do objects/figures get higher ratings, and what about the children drawings that draw eyes on frech fries etc ie mix the two?
% Unsurprisingly, the expert raters' scores had a higher standard deviation than the automated scores.

We visualize the drawings that received the highest scores in Figure \ref{figure5}. Interestingly, there are no overlaps in the top three highest scored drawings by expert and automated scores in any subgroup. This suggests that experts and the automated tools use very different strategies to rate creativity. 
% Drawings in Figure \ref{figure5} seem to suggest that the automated tools prefer drawings with more ink whereas the experts prefer drawings that either display interesting themes or use the base shape in interesting ways. 
We test this more formally in the next section.

% \subsubsection{Which group scores the maximum}
%     Avg, distribution of scores
% \subsubsection{Example high/low creative images}

\subsection{3. What do human experts and automated tools value when rating visual creativity?}
We use linear mixed effects regression to test the role of our style and content measures in explaining variance in mean creativity scores by experts and automated tools. We perform stratified, 3-fold cross validation and report $R^2_{test}$ and Speaman correlation $Cor_{test}$ metrics for the models with the lowest BICs (Table \ref{table2a}). We find that the expert score depended on originality, in both image and text space---\textit{i.e.} on how far the drawing is from the stimulus in image space ($dist\_from\_stim$), and on how similar the caption is to its ten nearest neighbours in text space ($10NN\_text$). Importantly, their scores also depended on how seamlessly the drawing incorporated the stimulus ($used\_stim$), and whether the drawing was easy to interpret ($hard\_to\_interpret$), which can be considered as evaluations of drawing utility. On the other hand, the automated score is largely influenced by the amount of ink ($ink\_density$) and originality measures operating purely largely in image space ($10NN\_image$), ignoring the multi-modal nature of creativity. 

We see this dissociation clearly in Table \ref{table2b}, which reports the regression coefficients for a combined linear model with all the predictors from the individual best models (VIFs are under 5). We see that the predictors useful for explaining experts scores, namely $used\_stim$, $hard\_to\_interpret$ and $10NN\_text$ are not effective predictors of mean automated score, while $ink\_density$ and $10NN\_image$ are not effective predictors of mean expert score. $Dist\_from\_stim$ was the only predictor contributing significantly to both models. 

% \subsection{4. How to Make AI drawings more similar to human drawings?}

\section{Discussion}
\vspace{-1mm}
Our work develops a novel dataset and a computational framework based on \textit{style} and \textit{content} to study creative drawings of children (across two age groups), adults and AI (across three prompts). 
% Style measures describe the visual and structural attributes capturing \textit{how} the drawing was made. Content measures focus on semantic and conceptual information, capturing \textit{what} the drawing is trying to express. Together, these measures provide a comprehensive framework for assessing both the execution and conceptualisation of a drawing.
This framework was useful in studying differences in drawings across the groups---children's drawings had more visual components, and depicted imaginative themes \cite{latham2018children}. Adult drawings displayed more conceptual diversity, and AI drawings had the highest ink density. We saw a striking increase in flexibility from pre-schematic to schematic children \cite{spensley1999development}. Also, content-driven prompting improved creativity in AI \cite{hao2024optimizing}. Unlike most textual creative tasks \cite{bellemare2024divergent, marco2024small}, Dall-e drawings still differ greatly from human outputs. Methods beyond prompting, for example agentic interactions \cite{vinker2024sketchagent} or fine-tuning could help align AI drawings more with humans'~\cite{liang2024rich}.
% The style-content framework seems to be useful in identifying and extracting differences across intelligent systems. 
 
% In verbal creativity, we can also split metrics up along these dimensions. Elaboration, i.e. num words, analogous to ink density. Given influence of ink density we hypothesize it is important to check for prediction of num words in verbal creativity. The question also arises to what extent style and content should be orthogonal.

Our framework also helps understand creativity and its measurement. Interestingly, which group was most creative depended on the evaluator. We confirm a self-bias where human experts preferred human drawings, and automated models preferred AI drawings \cite{magni2024humans}. As with textual creativity, both the automated methods and expert ratings value originality (distance from stimulus) \cite{kenett2019semdis} and novelty (distance from other responses) \cite{runco2012uniqueness}. But, the automated tools preferred ink density while expert scores valued utility. This highlights an important shortcoming of automated tools which lack understanding of underlying concepts and are therefore unable to incorporate the utility (effectiveness) dimension of creativity evaluation. However, this reflects positively on our search for domain-agnostic determinants of creativity---in line with past work, originality and utility interacted to predict creativity \cite{diedrich2015creative}. 
%\textit{Style} and \textit{content} have parallels with domain-specific and domain-general creativity respectively \cite{plucker2004creativity}, and our framework contributes to the debate on how to disentangle them \cite{stevenson2021mtc}. 

Despite challenges in scale unification across intelligent agents and manual curation of features, our framework, decomposing creativity into \textit{what} and \textit{how}, is general enough to extend beyond the visual domain to other forms of creative expression.

\bibliographystyle{apacite}
\setlength{\bibleftmargin}{.125in}
\setlength{\bibindent}{-\bibleftmargin}
\bibliography{references}

% %1. fMRI, visual creativity: https://onlinelibrary.wiley.com/doi/full/10.1002/brb3.540
% %2. co-creativity in drawing: https://dl.acm.org/doi/abs/10.1145/3563657.3595977?casa_token=s8V2WKJ-7LMAAAAA:Hdgzfr2hMOVnhYGmpMMexKTEdITjD_8TGVktrNZM6dgd3rKq5oiTkOvPGYFBj_WCmDYoAJpATIqoTm8
% %3. Human vs AI art perception: https://link.springer.com/article/10.1186/s41235-023-00499-6
% %4. Sketching partner AI: https://dl.acm.org/doi/abs/10.1145/3377325.3377522

\end{document}